# On a Class of Bias-Amplifying Variables that Endanger Effect Estimates

Judea Pearl
Cognitive Systems Laboratory
Computer Science Department
University of California, Los Angeles, CA 90024 USA
*judea@cs.ucla.edu*

## Abstract

This note deals with a class of variables that, if conditioned on, tends to amplify confounding bias in the analysis of causal effects. This class, independently discovered by Bhattacharya and Vogt (2007) and Wooldridge (2009), includes instrumental variables and variables that have greater influence on treatment selection than on the outcome. We offer a simple derivation and an intuitive explanation of this phenomenon and then extend the analysis to non linear models. We show that:

1. the bias-amplifying potential of instrumental variables extends over to non-linear models, though not as sweepingly as in linear models;

2. in non-linear models, conditioning on instrumental variables may introduce new bias where none existed before;

3. in both linear and non-linear models, instrumental variables have no effect on selection-induced bias.

## 1 INTRODUCTION

The common method of reducing confounding bias in the analysis of causal effects is to *adjust* for a set of variables judged to be "confounders," that is, variables capable of producing spurious associations between treatment and outcome, not attributable to their causal dependence. It is well known that a sufficient condition for the elimination of confounding bias is that the set of adjusted variable be "admissible," namely, that it satisfies the back-door criterion (Pearl, 1993, 2009a).[1] Moreover, if the chosen set is not admissible, adjustment for certain type of covariates may actually increase bias (Pearl, 1995, 2009a,c; Greenland et al., 1999; Heckman and Navarro-Lozano, 2004; Schisterman et al., 2009).[2] Such covariates include: (1) colliders, (2) intermediate variables on the causal pathways between treatment and outcome, and (3) descendants of such intermediaries (Weinberg, 1993; Pearl, 2009a, pp. 339–40).

Recently, Bhattacharya and Vogt (2007) and Wooldridge (2009) have identified a new class of covariates which, although not introducing new bias, tend to amplify bias if such exists.[3] They have shown that, in linear systems, conditioning on an instrumental variable (IV)[4] invariably causes an increase in confounding bias if such exists. IV measurements should therefore be discarded.

This result is far from obvious. First, an instrumental variable meets all the statistical properties that one normally associates with a confounder:

1. It is a pre-treatment variable, so, it is certainly not affected by the treatment, nor does it interfere with the causal pathways from treatment to outcome.

---

[1] Admissible sets are sometimes referred to as "sufficient" (Greenland et al., 1999) and, upon adjustment, the treatment is said to become "ignorable" (Rosenbaum and Rubin, 1983) or "unconfounded" (Pearl, 2009a). The "bias" considered here is bias of identification, not estimation, and is sometimes referred to as "inconsistency."

[2] The same applies to stratifying on these variables, using them for matching, using them as predictors in regression, or including them in inverse probability weighting or in the "propensity score" function (Rosenbaum and Rubin, 1983). The asymptotic equivalence of these methods is demonstrated in Pearl (2009a, pp. 348–51).

[3] A previous version of this paper attributed the discovery to Wooldridge (2009), I have subsequently become aware of the Working Paper by Bhattacharya and Vogt (2007).

[4] Roughly speaking, an instrumental variable is a variable associated with treatment but not with other factors that may affect outcome when treatment is fixed (by intervention). See (Pearl, 2009a, p. 248) for formal definitions and Fig. 1 for graphical representation.

2. It is related to treatment

3. It is related to outcome

4. It is related to outcome conditioned on treatment.

Thus, an IV seems to behave just like an ordinary confounder that begs to be controlled. Although the analysis of "confounding-equivalence" (Pearl, 2009a, pp. 345–6; Pearl and Paz, 2010) identifies IV's as bias modifiers, i.e., capable of increasing or decreasing bias, the idea that the modification always goes in the wrong direction (i.e., increased bias), is rather surprising, and calls for further analysis.

Second, the phenomenon stands contrary to current practices of covariate selection, such as those employed in propensity score methods. The prevailing practice is dominated by the belief that adding more covariates to the analysis can cause no harm (Rosenbaum, 2002, p. 76), especially covariates that are powerful predictors of the "treatment assignment." For example, a popular tutorial article by D'Agostino, Jr. (1998) states:

> "...if one has the ability to measure many of the covariates that are believed to be related to the treatment assignment, then one can be fairly confident that approximately unbiased estimates for the treatment effect can be obtained." p. 2267.

Rubin (2009) has further reinforced this attitude by stating that "To avoid conditioning on some observed covariates... is distinctly frequentist and non-scientific ad hockery."

This attitude relieves investigators from thinking about cause-effect relationships in the problem but is based on false premises (Pearl, 2009b) and leads, as noted by Bhattacharya and Vogt (2007), to wrong practical advice. For example, a major 2007 paper in the *Journal of the American Medical Association* advises investigators to include variables that are predictive of treatment assignment without regard to whether they are predictive of outcome (D'Agostino, Jr. and D'Agostino, Sr., 2007). The findings of this paper prove the opposite; disregarding the outcome leads to unwanted consequences. Whenever some residual bias persists, the more accurately one predicts the "treatment assignment" the higher the bias. More generally, we will show that the prevailing over-emphasis on modeling the "treatment assignment" while disregarding the outcome (Rubin, 2002) is misguided; modeling the "outcome mechanism" is in fact a much safer strategy for estimating causal effects in observational studies.

In this paper we derive (Section 2) and explain (Section 3) the bias-amplification phenomenon in a structural model setting and then extend the analysis in three directions. First (Section 4), we quantify the condition under which a confounding variable ceases to act as a bias reducer and instead becomes a bias amplifier. Second (Section 5), we consider non linear systems and show that the bias-amplification potential of instrumental variables extends over to non-linear models, though not as pervasively as in linear models; there are cases where conditioning on an IV reduces bias and, moreover, conditioning on instrumental variables may introduce new bias where none existed before. Finally (Section 6), we examine the effect of instrumental variables on selection bias created by preferential exclusion of units from the study (as in case-controlled studies,) and show that, in general, conditioning on an IV has no effect on such bias, unless the exclusion depends on factors that cause the treatment. The conclusion section summarizes the main findings of the paper and assesses their methodological ramifications in view of the on going conflict between the "experimentalist" and "structural" approaches to causal inference.

## 2 ANALYSIS

We will first consider a linear structural model, given in Fig. 1 where $X$ represents treatment, $Y$ is the outcome of interest, $U$ is an unobserved confounder, and $Z$ is an instrumental variable with respect to the causal effect of $X$ on $Y$. For simplicity, all variables will be assumed to be zero-mean and unit-variance. We will show that conditioning on $Z$ is harmful, for it increases the bias in the estimate of $c_0$.

We need to compare three quantities:

$$A_1 = \frac{\partial}{\partial x} E(Y|do(x))$$
$$A_2 = \frac{\partial}{\partial x} E(Y|x)$$
$$A_3 = \frac{\partial}{\partial x} E(Y|x, z)$$

$A_1$ is the incremental causal effect[5] of $X$ on $Y$, $A_2$ is the (unconditional) incremental dependence of $Y$ on $X$, given by the regression coefficient of $Y$ on $X$, and $A_3$ is the incremental conditional dependence of $Y$ on $X$, given by the coefficient of $X$ in the regression of $Y$

---

[5]Readers versed in potential-outcome notation can identify $E(Y|do(x))$ with the counterfactual expression $E(Y_x)$; both are given by the conditional expectation $E(Y|x)$ in a modified structural model, with $c_1$ and $c_3$ set to zero (see (Pearl, 2009a)).

on $X$ and $Z$. $A_2$ is often referred to as the "crude," or "naive" estimate of $A_1 = c_0$.

The unadjusted bias is given by the difference

$$B_0 = A_2 - A_1$$

while the bias after conditioning on $Z = z$ is given by

$$B_z = A_3 - A_1$$

Our task is to compare the magnitudes of $B_0$ and $B_z$ under various assumptions about the data-generating model.

For the model of Fig. 1, we have:

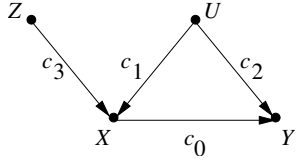

Figure 1: Linear model with instrumental variable $Z$ and confounder $U$.

$$A_1 = c_0 \qquad (1)$$
$$A_2 = c_0 + c_1 c_2 \qquad (2)$$
$$\begin{aligned} A_3 &= \frac{\partial}{\partial x} E(Y|x,z) = \frac{\partial}{\partial x} \sum_u E(Y|x,z,u) P(u|x,z) \\ &= \frac{\partial}{\partial x} \sum_u E(Y|x,u) P(u|x,z) \\ &= \frac{\partial}{\partial x} \sum_u (c_0 x + c_2 u) P(u|x,z) \\ &= c_0 + c_2 \frac{\partial}{\partial x} E(U|x,z) \end{aligned} \qquad (3)$$

Eq. (1) follows from the definition of $A_1$, Eq. (2) follows from Wright's rule of path analysis, and Eq. (3) was derived above using the structural assumption $E(Y|x,z,u) = c_0 x + c_2 u$.

Our next step is to evaluate the right hand side of Eq. (3) in terms of the structural coefficients $c_3$ and $c_1$. We start with the regression of $U$ on $X$ and $Z$:

$$u = \beta x + \alpha z + \epsilon \qquad (4)$$

where $\epsilon$ is the residual error and $\beta$ and $\alpha$ are chosen so as to minimize the square error $E[u - \beta x - \alpha z]^2$. It is well known that this minimization is achieved when $cov(\epsilon, x) = cov(\epsilon, z) = 0$ and then, if $E(U|x,z)$ is linear in $x$ and $z$, we have $E(U|x,z) = \beta x + \alpha z$. Therefore, to evaluate (3) we need to express $\beta$ and $\alpha$ in terms of $c_3$ and $c_1$. Multiplying (4) by $Z$ and $X$, taking expectations, and invoking the instrumental assumption $cov(Z, U) = 0$, we obtain:

$$\begin{aligned} E(UZ) &= \phantom{c_1} 0 \phantom{xx} = \beta E(XZ) + \alpha E(Z^2) = \beta c_3 + \alpha \\ E(UX) &= \phantom{0x} c_1 \phantom{x} = \beta E(X^2) + \alpha E(XZ) = \beta + \alpha c_3 \end{aligned}$$

yielding:

$$\beta = \frac{c_1}{1 - c_3^2} \qquad \alpha = -\frac{c_1 c_3}{1 - c_3^2} \qquad (5)$$

Substituting (5) in (3) enables us to evaluate $A_3$, giving:

$$A_3 = c_0 + c_2 \frac{\partial}{\partial x}(\beta x + \alpha z) = c_0 + \frac{c_2 c_1}{1 - c_3^2} \qquad (6)$$

We now have $A_1, A_2$, and $A_3$ evaluated, from which we can compute the biases $B_0$ and $B_z$, giving

$$B_z = \frac{c_2 c_1}{1 - c_3^2}, \quad B_0 = c_1 c_2, \quad B_z = \frac{B_0}{1 - c_3^2} \qquad (7)$$

Clearly, $|B_z| \geq |B_0|$ regardless of the signs of $c_1$ and $c_2$ with strict inequality holding whenever $|B_0| > 0$ and $|c_3| > 0$. The same result holds when $U$ is a vector of confounding variables. Thus, conditioning on $Z$ amplifies the unconditioned bias by a factor $\frac{1}{1-c_3^2}$.

## 3 INTUITION

For intuitive understanding of this phenomenon, consider the transition from $X = 0$ to $X = 1$, assuming a simplified equation $X = U + cZ$. By conditioning on $Z = z$, we are comparing units for which $U + cz = 0$ with those for which $U + cz = 1$. The mean difference in $U$ is of course unity

$$E(U|X=1, Z=z) - E(U|X=0, Z=z) = 1$$

and this difference will be transmitted to $Y$ and translate into the bias

$$E(Y|X=1, Z=z) - E(Y|X=0, Z=z) - c_0 = c_2$$

If, on the other hand, we do not condition on $Z$ but let it vary freely, the variation in $Z$ will "absorb," or "account for" part of the change in $X$, and only part of that change will be transmitted through $E(U)$ onto $E(Y)$. This can be seen quite clearly if we imagine that $U$ and $Z$ are uniformly distributed over the unit square. For any stratum $X = x$, $0 \leq x \leq 1$, $U$ will be uniformly distributed from 0 to $x$, with mean $E(U|X=x) = x/2$. Therefore, the mean difference (in $U$) between units for which $X = U + cZ = 0$ and those for which $U + cZ = x$ becomes

$$E(U|X=x) - E(U|X=0) = x/2 - 0 = x/2$$

which is merely half of the difference in the conditional expectation

$$E(U|X = x, Z = z) - E(U|X = 0, Z = z) = x$$

More generally, for any joint distribution of $U$ and $Z$, we can write:

$$\begin{aligned} E(U|x) &= \Sigma_z E(U|x,z)P(z|x) \\ &= \Sigma_z (x - cz)P(z|x) \\ &= x - cE(Z|x) \end{aligned}$$

from which we obtain

$$\begin{aligned} E(U|X = 1) &- E(U|X = 0) \\ &= 1 - c(E(Z|X = 1) - E(Z|X = 0)). \end{aligned}$$

The second term on the right is always positive, regardless of whether $X$ and $Z$ are positively or negatively correlated (respectively, $c > 0$ or $c < 0$).[6] We conclude therefore that mean difference in $U$ diminishes when we refrain from conditioning on $Z$:

$$\begin{aligned} E(U|X = 1) &- E(U|X = 0) \\ &< E(U|X = 1, Z = z) - E(U|X = 0, Z = z) \\ &= 1 \end{aligned}$$

and the bias transmitted onto $Y$ will likewise be reduced:

$$\begin{aligned} E(Y|X = 1) &- E(Y|X = 0) \\ &< E(Y|X = 1, Z = z) - E(Y|X = 0, Z = z). \end{aligned}$$

The last remark worth noting in this section is that an IV may easily exhibit Simpson's reversal, also called "suppression effect" or "Simpson's Paradox" (see Pearl (2009a) for survey and analysis). This can be seen from Eqs. (3) and (6); if $c_0$ and $c_1 c_2$ have opposite signs, $A_2$ and $A_3$ may be of opposite signs as well. This means that the association between $X$ and $Y$ in each stratum of $Z$ may be of opposite sign to the crude association, taken over all strata of $Z$. While such reversal is usually attributed to $Z$ being a confounder, the analysis above shows that the reversal can just as easily be generated by an IV. Indeed, this follows from the fact that, if $U$ is unobserved and $X$ continuous, the IV model of Fig. 1 is statistically indistinguishable from a model in which $Z$ is a cause of both $X$ and $Y$ (as in Fig. 2); every joint distribution $P(x, y, z)$ compatible with one model is also compatible with the other (Pearl, 2009a, pp. 274–5).

---

[6]More formally, we have:

$$\begin{aligned} E(U|X = x)/x &= Cov(XU)/Var(X) \\ &= Cov[U(U + cZ)/Var[(U + cZ)^2] \\ &= Var(U)/(Var(U) + c^2 Var(Z)) < 1 \end{aligned}$$

Thus, only a fraction of the unit change in $X$ will be transported over to $U$, and then to $Y$.

## 4 THE LINE BETWEEN INSTRUMENTS AND CONFOUNDERS

We now extend this result in three directions. First we relax the assumption that $Z$ is a "perfect" instrumental variable by allowing it to influence $Y$ directly as shown in Fig. 2. We ask for the relative values of $c_3$ and $c_4$ that would turn $Z$ from a bias-amplifier to a bias-reducer. Repeating the derivation of (6) with $E(Y|x, z, u) = c_0 x + c_4 z + c_2 u$ leaves $A_3$ and $B_z$ the same, but changes $A_2$ to read

$$A_2 = \frac{\partial}{\partial x} E(Y|x) = c_0 + c_2 c_1 + c_3 c_4 \qquad (8)$$

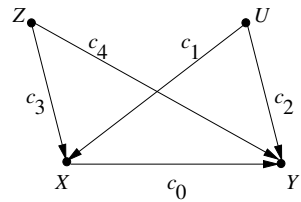

Figure 2: Model in which $Z$ is both a confounder and an (imperfect) instrumental variable relative to $X, Y$.

We see that $Z$ becomes a bias-reducer when

$$B_z \leq B_0 = c_2 c_1 + c_3 c_4$$

or

$$\frac{c_4}{c_3} \geq \frac{c_2 c_1}{1 - c_3^2} \qquad (9)$$

Thus, for $Z$ to become a bias-reducer, the effect of $Z$ on $Y$ must exceed its effect on $X$ by a factor $c_2 c_1 / 1 - c_3^2$. This may be a tall order to meet when $c_3$ is close to unity. Ironically, this means that the best predictors (of $X$) are also the most dangerous bias amplifiers.

This finding should be of concern to program evaluation researchers and propensity score analysts. Pretreatment covariates should be chosen for control or for propensity score matching, not because they are good predictors of treatment or outcome, but because they are deemed likely to reduce bias, when such is suspected. The analysis of this section shows that being a good predictor of treatment assignment compromises the bias-reducing potential of a covariate, for it tends to amplify bias due to other, uncontrolled confounders. One would do better therefore to rank order covariates based on their importance with respect to the outcome variable, a strategy advocated by Brookhart et al. (2006, 2010); Austin et al. (2007) and Hill (2008) which, in light of Eq. (9), deserves a general endorsement.

# 5 A GLIMPSE AT NON-LINEAR SYSTEMS

The second extension deals with the question of whether the amplification phenomenon described in (7), which prevails over all linear models regardless of parameter values extends over to non-linear models as well. We will show that, although the phenomenon persists in non-linear model, it is not as pervasive – there are non-linear models for which $|B_0| > |B_z|$.

Consider the model of Fig. 1, in which the equation determining $X$ remains the same,

$$X = c_3 z + c_1 u + \epsilon' \tag{9a}$$

but the one for $Y$ becomes non-linear in $X$:

$$Y = f(x) + u g(x) + \epsilon'' \tag{9b}$$

This yields

$$\begin{aligned} E(Y|x,z) &= f(x) + g(x) E(U|x,z) \\ &= f(x) + g(x)(\beta x + \alpha z) \\ &= f(x) + \beta g(x)(x - c_3 z) \end{aligned}$$

with $\beta$ and $\alpha$ given in (5); and, from (9a) and (9b)

$$\begin{aligned} E(Y|x) &= f(x) + g(x) E(U|x) \\ &= f(x) + g(x) c_1 x \end{aligned}$$

Consequently, $A_1, A_2,$ and $A_3$ evaluate to

$$A_1 = \frac{\partial}{\partial x} E(Y|do(x)) = \frac{\partial}{\partial x} E(f(x) + u g(x)) = f'(x)$$

$$A_2 = \frac{\partial}{\partial x} E(Y|x) = f'(x) + c_1(x g'(x) + g(x)) \tag{10}$$

$$A_3 = \frac{\partial}{\partial x} E(Y|x,z) = f'(x) + \beta(x g'(x) + g(x) - c_3 g'(x) z) \tag{11}$$

and the two bias measures become

$$B_0 = c_1(x g'(x) + g(x))$$
$$B_z = \frac{1}{1 - c_3^2}[c_1(x g'(x) + g(x) - c_3 g'(x) z)] \tag{12}$$
$$= \frac{1}{1 - c_3^2}(B_0 - c_1 c_3 g'(x) z)$$

Clearly, if $B_0 \geq 0$ and $c_1 c_3 g'(x) z > 0$, we can get $|B_z| < |B_0|$. This means that conditioning on $Z$ may reduce confounding bias, even though $Z$ is a perfect instrument and both $Y$ and $X$ are linear in $U$. Note that, owed to the non-linearity of $Y(x,u)$, the conditional bias depends on the value of $Z$ and, moreover, for $Z = 0$ we obtain the same bias amplification as in the linear case (Eq. (7)).

Equation (12) also shows that conditioning on $Z$ can introduce bias where none exists. This occurs when $c_1 > 0$ and $g(x) = A/x$, a condition that yields $B_0 = 0$ and $B_z > 0$. This potential of instrumental variables to produce new bias is suppressed in linear systems, as seen in Eq. (7), but is unleashed in non-linear systems. Still, this can only occur when $Z$ and $Y$ are dependent given $X$ (see (Pearl and Paz, 2010)); it will not occur therefore in situations where the zero bias condition $B_0 = 0$ is *structural*, that is, where one of the structural equations $x = h(z,u)$ or $y = h'(x,u)$ is trivial in its $u$ argument.[7]

# 6 THE RESILIENCE OF SELECTION BIAS

Our final extension concerns the effect of instrumental variables on selection bias, that is, bias induced by preferential selection of units for data analysis which is often governed by unknown factors including treatment, outcome and their consequences. Case control studies are particularly susceptible to such bias, e.g., when the outcome is a disease or complication that warrants reporting (see (Glymour and Greenland, 2008, pp. 111–37; Robins, 2001; Hernán et al., 2004)).

To illuminate the nature of this bias, consider the linear model of Fig. 3 in which $S$ is a variable affected by both $X$ and $Y$, indicating entry into the data pool. Such preferential selection to the data pool amounts to conditioning on $S$, and creates spurious association between $X$ and $Y$ through two mechanisms. First, conditioning on the collider $S$ induces spurious association between its parents, $X$ and $Y$. Second, $S$ is also a descendant of a "virtual collider" $Y$, whose parents are $X$ and the error term $U_Y$ (also called "omitted factors") which is always present, though often not shown.[8] The first mechanism is suppressed when $\beta_1$ is zero, while the latter is suppressed when $c_0$ is zero; both are suppressed when $\beta_2$ is zero.

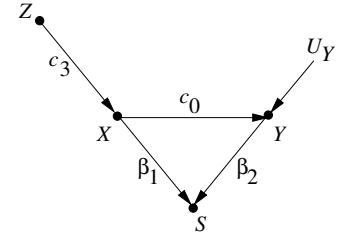

Figure 3: A model illustrating selection bias; conditioning on $S$ induces spurious associations between $X$ and $Y$.

---
[7]The condition $g(x) = A/x$ that gave rise to $B_0 = 0$ can be thought of as "unstable" (Pearl, 2009a, Ch. 2) for it depends critically on the exact form of the function $y = h'(x,u)$.

Writing

$$A_1 = \frac{\partial}{\partial x} E(Y|do(x)) = c_0 \qquad (13)$$

$$A_2 = \frac{\partial}{\partial x} E(Y|x,s) \qquad (14)$$

$$A_3 = \frac{\partial}{\partial x} E(Y|x,z,s) \qquad (15)$$

the bias due to conditioning on $S$, $A_2 - A_1$, can be calculated through the usual method of expectations (as in Eqs. (3)-(6)) and yields:

$$B_0 = A_2 - A_1 = \frac{-\beta_2(1-c_0^2)(\beta_1+c_0\beta_2)}{1-(\beta_1+c_0\beta_2)^2} \qquad (16)$$

We see that $B_0$ can be substantial and it vanishes if and only if one of the following conditions holds:

$$\beta_2 = 0, \quad c_0^2 = 1 \text{ or } \beta_1 = -c_0\beta_2.$$

In view of the amplification effect of IV's on confounding bias, one may be tempted to surmise that a similar effect can be expected vis-à-vis selection bias. This however is not the case. Conditioning on $Z$ has no effect whatsoever on selection-induced bias, formally,

$$A_3 = \frac{\partial}{\partial x} E(Y|x,s,z) = \frac{\partial}{\partial x} E(Y|x,s) = A_2 \qquad (17)$$

This equality can be derived, of course, from the parametric model of Fig. 3, going through the necessary (yet painstaking) steps of algebraic manipulations. However, the validity of this equality is much broader, for it holds in non-linear systems as well. This can be seen immediately from the structure of the diagram of Fig. 3, which asserts that, regardless of the functional relationships between the variables in the diagram, $Y$ and $Z$ are independent given $X$ and $S$,[9] which entails Eq. (17).

We thus conclude that selection bias differs fundamentally from confounding bias in that the former, as distinct from the latter, is insensitive to conditioning on an IV. This distinction can be used in practice to detect the presence of confounding bias. If one has a solid theoretical basis to believe that a variable $Z$ is a valid instrument relative to the effect of $X$ on $Y$, and if data shows that the association between $X$ and $Y$ changes upon conditioning on $Z$, chances are the study is marred by confounding bias, and remedial steps are necessary, possibly through covariate control. Conversely, if no such changes can be detected in the data, chances are no confounding bias exists, though the study can still be contaminated with selection-induced bias, resistant to covariate control.[10]

---

[8]See (Pearl, 2009a, pp. 339–41) for further explanation of this bias mechanism, which seems to have escaped the taxonomies in (Hernán et al., 2004) and (Schisterman et al., 2009).

[9]This is verified through the *d*-separation rule (Pearl, 2009a, pp. 335–7), which identifies the conditional independencies implied by a system of non-linear structural equations.

The question naturally arises whether IV-sensitivity can also be used to assess the relative magnitude of the biases associated with two or more effect estimates, each produced by a different method of covariate control. The answer is trivially affirmative in linear models, since all effects are identifiable in the IV model of Fig. 1. In nonlinear models, where causal effects are nonidentifiable, the IV-sensitivity can serve to rank estimates, but the utility of such ranking is limited; low sensitivity, hence low bias, may as well be assessed indirectly, by measuring the degree to which $Y$ and $Z$ are dependent given $X$.

It is important to keep in mind though that the selection bias analyzed above was "pure," in the sense of inducing no confounding component. In general, if the reasons for excluding units from the study data involves ancestors of $X$, confounding bias may also be induced, as shown in Fig. 4. $S_3$ induces "pure" con-founding bias that would be amplified by conditioning on $Z$, while $S_2$ induces a "pure" selection bias that will be impervious to conditioning on $Z$. $S_1$ induces both selection and confounding bias through to the path $X - U_1 - S_1 - Y$; the latter will be sensitive to $Z$. Conditioning on $U_2$ will eliminate the entire bias induced by $S_2$, while conditioning on $U_1$ will eliminate only the confounding part of the bias induced by $S_1$; the selection-induced part, due to the association created between $X$ and $U_Y$, cannot be eliminated by any method. $S_3$ induces purely confounding bias, which can be eliminated by conditioning on either $U_1$ or $U_2$.

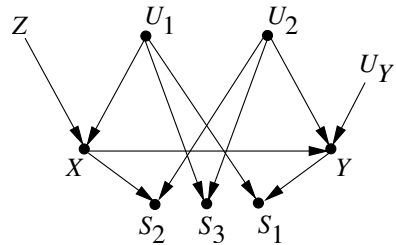

Figure 4: Model in which conditioning on $S_2$ induces selection bias, conditioning on $S_3$ induces confounding bias, and conditioning on $S_1$ induces both.

---

[10]I use the cautionary term "chances are" to allow for the rare possibilities that $Z$ introduces its own bias (see footnote 7) or that the bias will not be changed by $Z$. These possibilities are not structural, in the sense that they require fine tuning of the functional relationships in the model.

Formally, the distinction between confounding and selection bias can be articulated thus: Confounding bias is any $X - Y$ association that is attributable to paths traversing ancestors of $X$ (i.e., factors affecting treatment.) Operationally, the distinction may refer to experimental paradigm: Confounding bias is any $X - Y$ association that can be eliminated by randomization. The theory of causal diagrams (Pearl, 2009a; Spirtes et al., 2000) establishes the equivalence of the two criteria.

# 7  CONCLUSIONS

We have examined the effect of instrumental variables on various types of bias. We first showed that, while in linear systems conditioning on an IV always amplifies confounding bias (if such exists), bias in nonlinear systems may be amplified as well as attenuated. In some cases an IV may introduce new bias where none exists. We further examined the effect of IV's on selection-induced bias and showed that no such effect exists as long as the bias contains no confounding component. A formal criterion for distinguishing the two types of bias sources was introduced, and the possibility of using IV-sensitivity as a diagnostic tool for bias detection was suggested.

From a practical viewpoint, the immediate implication of this analysis is that covariates should be chosen based on their importance with respect to the outcome, rather than the treatment. Those that have meager effects on the outcome mechanism and strong effects on the treatment should be discarded, to prevent bias amplification, while those that have strong effects on the outcome mechanism should be retained and serve as predictors in the propensity score. In the absence of unobserved confounders, a propensity score constructed from the direct causes of the outcome gives the same effect estimate (asymptotically) as one constructed from the direct causes of the treatment (Pearl, 2009a, pp. 348–52); the former, however, is safer in the presence of residual unobserved confounders.

The phenomenon illuminated in this paper also has a profound methodological significance on the conflict between the so-called "structural" and "experimentalist" camps in causal analysis. The former requires that every causal analysis commences with an explicit representation of the available causal assumptions behind the study, be it in the form of structural equations or in their non-parametric version as causal diagrams (Pearl, 2010). The latter attempts to avoid such representation, fearing that causal assumptions, if made transparent, would be deemed indefensible.

The consequences of avoiding structural considerations can be seen through the ill-advised practices that the "experimentalist" approach has nourished in the past two decades. Unprincipled covariate selection is one such practice.

Guido Imbens, for example, one of the staunchest proponents of the "experimentalist" approach, recommends that variables be selected for the propensity score as follows: "After estimating all [univariate] logistic regressions we end up with the subset of covariates whose marginal correlation with the treatment indicator is relatively high. We orthogonalize the set of selected covariates, and use these to estimate the propensity score" (Hirano and Imbens, 2001). Reluctance to consider the causal structure of the problem and, in particular, whether candidate covariates affect the outcome may easily invite strong predictors that amplify, rather than diminish confounding bias.

Bhattacharya and Vogt (2007) attribute this attitude of the experimentalist approach to the traditional reluctance of statisticians to rely on theoretical or judgmental assumptions. Pearl (2009b) on the other hand attributes this reluctance to the deficiency of the potential-outcome notation which, lacking transparency, discourages the articulation of causal assumptions, even those defensible on scientific grounds. Whichever the case, there is no substitute to structural knowledge in causal analysis.

**Acknowledgments**

This note has benefited from discussions with Jeffrey Wooldridge, Jennifer Hill, Antonio Forcina, and Peter Austin, and was supported in parts by grants from NIH #1R01 LM009961-01, NSF #IIS-0914211, and ONR #N000-14-09-1-0665.